\documentclass[aps,10pt]{revtex4}
\usepackage{amsmath}
\usepackage{amssymb}
\usepackage{hyperref}
\hypersetup{colorlinks=true,linkcolor=blue,citecolor=red,anchorcolor=black}
 
\begin{document}

\title{Solution to the ghost problem in higher-derivative gravity}
\author{Philip D. Mannheim}
\affiliation{Department of Physics \\ University of Connecticut \\ Storrs, CT 06269, USA 
\\ philip.mannheim@uconn.edu}

\date{September 26 2021}

\begin{abstract}
With standard Einstein gravity not being renormalizable at the quantum level there is much interest in studying higher-derivative quantum gravity theories. Thus just as a Ricci-scalar-based action produces a propagator that behaves as a non-renormalizable $1/k^2$ at large $k^2$, an action based on the square of the Ricci scalar behaves as a renormalizable $1/k^4$ at large $k^2$. An action based on both the Ricci scalar and its square leads  to a renormalizable propagator of the generic Pauli-Villars form. However, given the form of the Hamiltonian and the propagator such theories are thought to be plagued by either energies that are unbounded from below or states of negative Dirac norm (the overlap of a ket with its Hermitian conjugate bra). But when one constructs the quantum Hilbert space one finds (Bender and Mannheim) that in fact neither of these problems is actually present. The Hamiltonian turns out to not be Hermitian but to instead have an antilinear $PT$ symmetry, and for this symmetry the needed inner product is the overlap of a ket with its $PT$ conjugate bra. And this inner product is positive definite. Moreover, for the pure $1/k^4$ propagator the Hamiltonian turns out to not be diagonalizable, and again there are no states of negative energy or of negative norm. Instead there are states of zero norm, non-standard but perfectly acceptable states that serve to maintain probability conservation. With the locally conformal invariant fourth-order derivative conformal gravity theory being in this category, it can be offered as a candidate theory of quantum  gravity that is renormalizable and unitary in four spacetime dimensions.

\end{abstract}

\maketitle

\section{Pauli-Villars regulator and the seventy year ghost problem}

With the prototype for the higher-derivative gravity studies initiated in \cite{Stelle1977} being the Pauli-Villars propagator, we analyze its generic features as they also apply to higher-derivative gravity itself.
In order to regularize the divergences  in loop graphs and also to implement gauge invariance Pauli and Villars \cite{Pauli1949} proposed that one replace the standard one-particle $1/(k^2-M^2)$ propagator by a two-particle second-order propagator
\begin{eqnarray}
D(k,2)=\frac{1}{k^2-M_1^2}-\frac{1}{k^2-M_2^2},
\label{1.1}
\end{eqnarray} 
so that a $1/k^2$ asymptotic ultraviolet behavior at large $k^2$ would be replaced by the more convergent $1/k^4$, with a quadratically divergent one loop scalar field graph then becoming only  logarithmically  divergent. 
While Pauli and Villars recognized this as a mathematical procedure, they did not want to rule out  that it might be physical.
As conceived, the two particles would be associated with the second-order derivative action 
\begin{eqnarray}
\quad I_{1}+I_{2}&=\tfrac{1}{2}\int d^4x\left[\partial_{\mu}\phi_1\partial^{\mu}\phi_1-M_1^2\phi_1^2-\lambda\phi_1^4\right]
+\tfrac{1}{2}\int d^4x\left[\partial_{\mu}\phi_2\partial^{\mu}\phi_2-M_2^2\phi_2^2-\lambda\phi_2^4\right].
\label{2.1}
\end{eqnarray}
Following the insertion of the very specific closure relation
\begin{eqnarray}
\sum |n_1\rangle\langle n_1|-\sum |n_2\rangle\langle n_2|=I
\label{2.3}
\end{eqnarray}
with its ghostlike relative minus sign into
\begin{eqnarray}
D(k,2)=\int d^4x e^{ik\cdot x}[\langle\Omega_1|T[\phi_1(x)\phi_1(0)]|\Omega_1\rangle+\langle\Omega_2|T[\phi_2(x)\phi_2(0)]|\Omega_2\rangle]
\label{2.3a}
\end{eqnarray}
we recover (\ref{1.1}). Thus ghosts could reduce asymptotic divergences, but at the price of loss of probability and loss of unitarity.

\section{The Pais-Uhlenbeck Oscillator}

To explore whether the Pauli-Villars regulator might be physical Pais and Uhlenbeck \cite{Pais1950}  replaced the two-field action by a one-field fourth-order derivative action 
\begin{eqnarray}
I_S&=&\tfrac{1}{2}\int d^4x\bigg{[}\partial_{\mu}\partial_{\nu}\phi\partial^{\mu}
\partial^{\nu}\phi-(M_1^2+M_2^2)\partial_{\mu}\phi\partial^{\mu}\phi
+M_1^2M_2^2\phi^2\bigg{]},
\label{2.4}
\end{eqnarray}
with a  fourth-order derivative equation of motion given by
\begin{eqnarray}
&&(\partial_t^2-\vec{\nabla}^2+M_1^2)(\partial_t^2-\vec{\nabla}^2+M_2^2)
\phi(x)=0,
\label{2.5}
\end{eqnarray}
and associated fourth-order propagator of the Pauli-Villars form
\begin{eqnarray}
D(k,4)=\frac{1}{(k^2-M_1^2)(k^2-M_2^2)}=\frac{1}{(M_1^2-M_2^2)}\left(\frac{1}{k^2-M_1^2}-\frac{1}{k^2-M_2^2}\right).
\label{2.6}
\end{eqnarray}
So is this theory also not unitary? It has the same propagator structure as the second-order $D(k,2)$. If we identify the fourth-order propagator as 
\begin{equation}
D(k,4)=\int d^4x e^{ik\cdot x}\langle \Omega|T[\phi(x)\phi(0)]|\Omega\rangle,
\label{2.5x}
\end{equation}
then the insertion of the  ghostlike closure relation given in (\ref{2.3}) into 
 $\langle \Omega|T[\phi(x)\phi(0)]|\Omega\rangle$ would generate the $D(k,4)$ propagator. So could anything be wrong with this reasoning?

\section{Why something must be wrong and why there must be an underlying conformal symmetry and a fourth-order derivative action}

Consider the Dirac action for a massless fermion coupled to a background geometry of the form
\begin{eqnarray}
I_{\rm D}=\int d^4x(-g)^{1/2}i\bar{\psi}\gamma^{c}V^{\mu}_c(\partial_{\mu}+\Gamma_{\mu})\psi, 
\label{5.3a}
\end{eqnarray}
where the $V^{\mu}_a$ are vierbeins, $\Gamma^{\lambda}_{\mu\nu}=(1/2)g^{\lambda\sigma}(\partial_{\mu}g_{\sigma \nu} +\partial_{\nu}g_{\sigma \mu}-\partial_{\sigma}g_{\mu \nu})$ is the Levi-Civita connection,   and $\Gamma_{\mu}=-(1/8)[\gamma_a,\gamma_b](V^b_{\nu}\partial_{\mu}V^{a\nu}+V^b_{\lambda}\Gamma^{\lambda}_{\nu\mu}V^{a\nu})$ is the spin connection that enables $I_{\rm D}$ to be locally Lorentz invariant. 
As constructed, and while not designed for this purpose, $\Gamma_{\mu}$ also enables $I_{\rm D}$ to be  locally conformal invariant under the local conformal transformation
$$V^{\mu}_a\rightarrow e^{-\alpha(x)}V^{\mu}_a(x),~~~\psi(x)\rightarrow e^{-3\alpha(x)/2}\psi(x),~~~g_{\mu\nu}(x)\rightarrow e^{2\alpha(x)} g_{\mu\nu}(x).$$
We thus get local conformal invariance for free. In fact, other than the double-well potential, the entire $SU(3)\times SU(2)\times U(1)$ standard model is locally conformal invariant. Thus if fermion masses are generated dynamically (see e.g. \cite{Mannheim2017} and references therein), then the entire standard model would be locally conformal invariant. 't Hooft \cite{tHooft2015b}  has also argued that there should be an underlying local conformal symmetry in nature. Thus for a generic $\psi\rightarrow e^{\alpha+i\beta}\psi$, gauging $\beta$ is Yang-Mills, gauging $\alpha$ is conformal gravity. Thus we can unify $SU(3)\times SU(2)\times U(1)$ with gravity through real and imaginary local phases.

But local conformal invariance requires that the action in the gravitational sector  be quadratic in the Ricci tensor and  scalar and be of the generic form given in (\ref{5.4}). To see this we set  the path integral $\int D[\psi]D[\bar{\psi}]\exp(iI_{\rm D})$ equal to $\exp(iI_{\rm EFF})$,  integrate on $\psi$ and $\bar{\psi}$,  and obtain an effective action with leading term  ('t Hooft \cite{tHooft2010a})
\begin{eqnarray}
I_{\rm EFF}&=&\int d^4x(-g)^{1/2}C\left[R_{\mu\nu}R^{\mu\nu}-\tfrac{1}{3}(R^{\alpha}_{\phantom{\alpha}\alpha})^2\right],
\label{5.4}
\end{eqnarray}
where $C$ is a log divergent constant. With $I_D$ being locally conformal invariant, $I_{\rm EFF}$ must be locally conformal invariant too, just as is in fact the case \cite{Mannheim2006}. Thus in the standard model itself we generate the fourth-order derivative conformal gravity action. So have we lost unitarity? No we could not have, since with the standard model being ghost free, and with the fermion path integral being equivalent to a one loop Feynman diagram, then  since we cannot change the signature of a Hilbert space in perturbation theory, conformal gravity must be ghost free too. And indeed if were not then the standard model would lose unitary when coupled to gravity.
Thus radiative corrections to $SU(3)\times SU(2)\times U(1)$ in an external gravitational field generate fourth-order conformal gravity whether we like it or not. So we have to deal with conformal gravity one way or the other.

\section{So where could things possibly go wrong?}

Starting from a c-number propagator we cannot determine the structure of the underlying q-number theory.  Thus we cannot identify the c-number $D(k,4)$ with the matrix element $\langle \Omega|T[\phi(x)\phi(0)]|\Omega\rangle$ before first constructing the quantum theory. We can of course construct c-numbers from q-numbers but not vice versa.
Thus just because $D(k,2)$ is associated with a loss of unitarity it does not mean that $D(k,4)$ is too. To find out we need to construct the quantum Hilbert space. 

\section{So where do things go wrong?}

As noted by Bender and Mannheim \cite{Bender2008a,Bender2008b}, the right eigenvacuum of the fourth-order quantum theory associated with the $D(k,4)$ propagator obeys $H|\Omega_R\rangle=0$.  The left eigenvacuum obeys $\langle \Omega_L|H=0$. In general the left eigenvacuum is only the Hermitian conjugate of the right eigenvacuum if the Hamiltonian is Hermitian. So is it Hermitian? This depends on boundary conditions. It turns out that the boundary conditions are such that the $D(k,4)$ Hamiltonian is not in fact Hermitian (we cannot integrate by parts and throw surface terms away). However, it instead is $CPT$ symmetric, and the left eigenvacuum $\langle \Omega_L|$ is the $CPT$ conjugate of $|\Omega_R\rangle$. In consequence the $D(k,4)$ propagator has to be identified with $\int d^4x e^{ik\cdot x}\langle \Omega_L|T[\phi(x)\phi(0)]|\Omega_R\rangle$, and this propagator is unitary \cite{Bender2008a,Bender2008b}, even as $\int d^4x e^{ik\cdot x}\langle \Omega_R|T[\phi(x)\phi(0)]|\Omega_R\rangle$ is not.

Since $C$ (charge conjugation) is separately conserved for gravity, $H$ is $PT$ ($P$ is parity, $T$ is time reversal) symmetric, and thus falls into the class of $PT$ symmetric theories initiated by Bender (see \cite{Bender2018} for a recent overview), with antilinear $CPT$ symmetry  being recognized as being more general for quantum theory than Hermiticity \cite{Mannheim2018a}. However, even though the Hamiltonian is not Hermitian, all the poles of $D(k,4)$ are on the real axis. So all energy eigenvalues are real, the hallmark of $PT$ theories. Hermiticity is only sufficient to give real eigenvalues, $CPT$ symmetry is necessary (see \cite{Mannheim2018a} and references therein). Also, since an overall minus sign does not affect the causal domain of Feynman contours, $D(k,4)$ is causal. Thus $D(k,4)$ is associated with $\langle \Omega_L|T[\phi(x)\phi(0)]|\Omega_R\rangle$, and the $\langle L|R\rangle$ inner product is positive definite (see \cite{Mannheim2018a,Mannheim2018c} and the discussion below). We cannot use  (\ref{2.3}) since the states  are not normalizable. So now to the proof.

\section{$PT$ symmetry or antilinear symmetry in general}

Consider the eigenvector equation 
\begin{eqnarray}
i\frac{\partial}{\partial t}|\psi(t)\rangle=H|\psi(t)\rangle=E|\psi(t)\rangle.
\label{H1b}
\end{eqnarray}
Replace the parameter $t$ by $-t$ and then multiply by some general antilinear operator $A$:
\begin{eqnarray}
i\frac{\partial}{\partial t}A|\psi(-t)\rangle=AHA^{-1}A|\psi(-t)\rangle=E^*A|\psi(-t)\rangle.
\label{H2b}
\end{eqnarray}
If $H$ has an antilinear symmetry so that $AHA^{-1}=H$, then  
\begin{eqnarray}
HA|\psi(-t)\rangle=E^*A|\psi(-t)\rangle.
\label{H2bb}
\end{eqnarray}
There are two possibilities: (1) (Wigner): Energies can be real and have eigenfunctions that obey $A|\psi(-t)\rangle=|\psi(t)\rangle$, 
(2) or energies can appear in complex conjugate pairs that have conjugate eigenfunctions ($|\psi(t)\rangle \sim \exp(-iEt)$ and $A|\psi(-t)\rangle\sim \exp(-iE^*t)$).
Thus with realization (1) we obtain real eigenvalues with antilinear symmetry. 

As to the converse, suppose we are given that the energy eigenvalues are real or appear in complex  conjugate pairs. In such a case not only would $E$ be an eigenvalue but $E^*$ would be too. Hence, we can set $HA|\psi(-t)\rangle=E^*A|\psi(-t)\rangle$ in (\ref{H2b}), and obtain
\begin{eqnarray}
(AHA^{-1}-H)A|\psi(-t)\rangle=0.
\label{H3b}
\end{eqnarray}
Then if the eigenstates of $H$ are complete, (\ref{H3b}) must hold for every eigenstate, to yield $AHA^{-1}=H$ as an operator identity, with $H$ thus having an antilinear symmetry. Antilinear symmetry is thus \textbf{necessary} for the reality of eigenvalues while Hermiticity is only \textbf{sufficient}.

For inner products we note that the Dirac inner product $\langle R(t)|e^{iH^{\dagger} t}e^{-iHt}|R(t)\rangle$ is only time independent if $H=H^{\dagger}$. However, the left-right inner product $\langle L(t)|e^{iH t}e^{-iHt}|R(t)\rangle$ is always time independent regardless of whether or not $H=H^{\dagger}$. This left-right inner product is also equal to the overlap of a ket with its antilinear symmetry conjugate (see e.g. \cite{Mannheim2018a,Mannheim2018c}). And in fact on very general grounds it has been shown \cite{Mannheim2018a}   that under only two requirements, namely invariance under the complex Lorentz group (the proper Lorentz group) and probability conservation the quantum Hamiltonian must be $CPT$ symmetric, being so without having any need to be Hermitian. Since $CPT$ defaults to $PT$ for non-relativistic quantum systems this puts the $PT$ symmetry program on a very secure theoretical basis. In the following then we shall look for a connection between higher-derivative gravity and the $PT$ program. So we start with the Pais-Uhlenbeck Oscillator model.

\section{Quantizing the Pais-Uhlenbeck Oscillator Model}

Since only time derivatives are relevant to quantization, we set $\omega_1=(\bar{k}^2+M_1^2)^{1/2}$, $\omega_2=(\bar{k}^2+M_2^2)^{1/2}$ and drop the spatial dependence. With $z$ replacing $\phi$ the $I_S$ action given in (\ref{2.4}) reduces to the Pais-Uhlenbeck ($PU$) action
\begin{eqnarray}
I_{\rm PU}=\tfrac{1}{2}\int dt\left[{\ddot z}^2-\left(\omega_1^2
+\omega_2^2\right){\dot z}^2+\omega_1^2\omega_2^2z^2\right].
\label{2.7}
\end{eqnarray}
This is a constrained action since with only $z$, $\dot{z}$ and $\ddot{z}$, there are too many canonical variables for one oscillator but not enough for two. So we set $x=\dot{z}$, and using the method of Ostrogradski \cite{Ostrogradski1850}, using Dirac Constraints \cite{Mannheim2005}, or a variation of a covariantized form of the action with respect to the metric, one obtains the two-oscillator $PU$ Hamiltonian 
\begin{eqnarray}
H_{\rm PU}=\tfrac{1}{2}p_x^2+p_zx+\tfrac{1}{2}\left(\omega_1^2+\omega_2^2 \right)x^2-\tfrac{1}{2}\omega_1^2\omega_2^2z^2, \quad [z,p_z]=i, \quad [x,p_x]=i.
\label{2.10}
\end{eqnarray}
Now there are no ghosts, but one now has to pay a different price: the  $-\tfrac{1}{2}\omega_1^2\omega_2^2z^2$  term in $H_{\rm PU}$ leads to an energy spectrum that is unbounded from below, the Ostrogradski  \cite{Ostrogradski1850} instability that is characteristic of higher-derivative theories. 

\section{Trading the energy instability for ghosts}

Work by Lee \cite{Lee1954}, K\"all\'en and Pauli \cite{Kallen1955}  and Heisenberg \cite{Heisenberg1957} reopened the ghost issue, and it was found that one could avoid negative energies in the $PU$ theory if one quantized the $PU$ theory with negative norm states.
Specifically, if  we make the standard substitutions
\begin{eqnarray}
z&=&a_1+a_1^{\dagger}+a_2+a_2^{\dagger},\quad p_z=i\omega_1\omega_2^2
(a_1-a_1^{\dagger})+i\omega_1^2\omega_2(a_2-a_2^{\dagger}),
\nonumber\\
x&=&-i\omega_1(a_1-a_1^{\dagger})-i\omega_2(a_2-a_2^{\dagger}),\quad
p_x=-\omega_1^2 (a_1+a_1^{\dagger})-\omega_2^2(a_2+a_2^{\dagger}),
\label{2.11}
\end{eqnarray}
we obtain a Hamiltonian and commutator algebra \cite{Mannheim2005} 
\begin{align}
H_{\rm PU}&=2(\omega_1^2-\omega_2^2)(\omega_1^2 a_1^{\dagger}
a_1-\omega_2^2a_2^{\dagger} a_2)+\tfrac{1}{2}(\omega_1+\omega_2),
\nonumber\\
[a_1,a_1^{\dagger}]&=\frac{1}{2\omega_1(\omega_1^2-\omega_2^2)},\quad [a_2,a_2^{\dagger}]=-\frac{1}{2\omega_2(\omega_1^2-\omega_2^2)},
\label{2.12}
\end{align}
and note that with  $\omega_1>\omega_2$ the $[a_2,a_2^{\dagger}]$ commutator is negative.  

Also,  we note that the $\omega_1=\omega_2$  limit is singular, an issue we will return below. However,  we note now that we cannot write the $\omega_1=\omega_2$ propagator (i.e., $M_1^2=M_2^2$) as 
\begin{equation}
\frac{1}{(k^2+i\epsilon)^2}=\lim_{M_1^2\rightarrow 0, M_2^2\rightarrow 0} \frac{1}{(M_1^2-M_2^2)}\left(\frac{1}{k^2-M_1^2+i\epsilon}-\frac{1}{k^2-M_2^2+i\epsilon}\right),
\label{2.12x}
\end{equation}
since the  $M_1^2\rightarrow 0$, $M^2_2\rightarrow 0$ limit is singular. However, we could write the propagator as  
\begin{equation}
\frac{1}{(k^2+i\epsilon)^2}= \lim_{M^2\rightarrow 0} \frac{d}{d M^2}\left(\frac{1}{k^2+M^2+i\epsilon}\right),
\label{2.12xy}
\end{equation}
a non-singular limit in which no negative sign appears. With this form we can anticipate that the pure fourth-order theory is unitary, just as was in fact proven in \cite{Bender2008a,Bender2008b}. We return to this point below.

There are two realizations of (\ref{2.12}). 
If we define the vacuum according to 
$$a_1|\Omega\rangle=0,~~~a_2|\Omega\rangle=0,$$ 
the energy spectrum is bounded from below, with $|\Omega\rangle$ being the ground state with energy $\left(\omega_1+\omega_2\right)/2$. But  the excited state
$a_2^\dag|\Omega\rangle$, which lies at energy $\omega_2$ above the ground state, has a Dirac norm $\langle\Omega|a_2a_2^\dag|\Omega\rangle$ that is negative. On the other hand if we define the vacuum according to 
$$a_1|\Omega\rangle=0,~~~a_2^{\dagger}|\Omega\rangle=0,$$
the $PU$ theory is now free of negative-norm states, but the energy spectrum is unbounded from below. (Negative energy states propagating forward in time.) However, for either realization  all the eigenvalues of $H_{\rm PU}$ are real. This will  be of importance in the following.

Thus the $PU$ theory suffers from one of two twin diseases, either negative norms or negative energies. Since defining the vacuum by setting $a_2|\Omega\rangle=0$ or by setting $a_2^{\dagger}|\Omega\rangle=0$ would correspond to working in two totally different Hilbert spaces, in no single Hilbert space does one have both diseases, though in either one there is still a seemingly irrefutable  problem.
The objective of this paper is not to find a way to cancel ghost states in the  $a_2|\Omega\rangle=0$ realization in which there are no negative energies, but to show that the reasoning that leads one to think that there is a ghost is faulty. Thus we do not get rid of the ghosts but instead \cite{Bender2008a,Bender2008b} show that they were never there to begin with. That this could in principle be possible is because a propagator such as $D(k,4)$ is a c-number, and from a c-number one cannot construct the underlying q-number Hilbert space. One can construct c-number matrix elements from q-number operators but not the other way round. Thus we need to find the relevant Hilbert space and need to find a relevant positive definite inner product that would replace the standard Dirac norm.

\section{$PT$ Symmetry to the rescue -- the Lee model}

In 1954 Lee introduced a model in which one could do coupling constant renormalization analytically.  However, the model had ghost states of negative norm and Lee, K\"all\'en and Pauli, and Heisenberg worked very hard on the issue. However, the problem remained unsolved until the work of  Bender, Brandt, Chen and Wang \cite{Bender2005}  no less than some fifty years later.  In the Lee model ghost states only appear for a certain range of values of the renormalized coupling constant of the model, and in that range the bare coupling constant is complex. In consequence, in that range  the theory is not a Hermitian theory, and one cannot use as norm or inner product  the overlap of a ket state with its Hermitian conjugate bra. However, in \cite{Bender2005} it was found that the theory has an antilinear $PT$ symmetry, and when one uses the $PT$ theory norm, viz. the overlap of a ket with its $PT$ conjugate, one finds that this norm is positive definite. Solving the Lee model ghost problem this way is a considerable triumph for $PT$ theory. Thus in general if one finds states of negative Dirac norm, it does not necessarily mean that the theory is not unitary. It could mean that one is in the wrong Hilbert space and that one is using the wrong inner product (the Dirac one), with a different one (the $PT$ one) being unitary. 

\section{Implications for the Pauli-Villars propagator and the $PU$ Oscillator}

On returning  to the $PU$ Hamiltonian given in (\ref{2.10}) we note that unlike the Lee model  there is no complex coupling constant that could save us. However, if we  set $p_z=-i\partial_z$, $p_x=-i\partial_x$, the Schr\"odinger equation takes the form
\begin{eqnarray}
\left[-\frac{1}{2}\frac{\partial^2}{\partial x^2}-ix\frac{\partial}{
\partial z}+\frac{1}{2}(\omega_1^2+\omega_2^2)x^2-\frac{1}{2}
\omega_1^2\omega_2^2z^2\right]\psi_n(z,x)=E_n\psi_n(z,x),
\label{2.13a}
\end{eqnarray}
with the lowest positive energy state with $E_0=(\omega_1+\omega_2)/2$ having eigenfunction \cite{Bender2008a} 
\begin{eqnarray}
\psi_0(z,x)={\rm exp}\left[\tfrac{1}{2}(\omega_1+\omega_2)\omega_1\omega_2
z^2+i\omega_1\omega_2zx-\tfrac{1}{2}(\omega_1+\omega_2)x^2\right].
\label{2.14a}
\end{eqnarray}
The state $\psi_0(z,x)$ diverges as $z\rightarrow \infty$ and is thus not normalizable. Consequently, the norm of the ground state 
$$\langle \Omega |\Omega\rangle=\int dzdx\langle\Omega |z,x\rangle\langle z,x|\Omega\rangle=\int dzdx\psi^*_0(z,x)\psi_0(z,x)$$  
is infinite too. Such lack of normalizability means that the closure relation 
given in (\ref{2.3}) could not hold as it presupposes normalizable states. However, rather than being a bad thing,  \textbf{it is the lack of normalizability that actually saves the theory.}

Specifically, to make the wave function $
\psi_0(z,x)$ normalizable we must continue $z$ into the complex plane. If we draw a letter $X$ in the complex $z$ plane, the wave function will be normalizable in a wedge (a so-called Stokes wedge -- actually a generic feature of $PT$ theories \cite{Bender2018}) that contains the north and south quadrants of the letter $X$, i.e. that contains the imaginary $z$ axis but not the real $z$ axis (which is in the east and west quadrants). To implement this, with $p_x=p$  we make the similarity transformations \cite{Bender2008a}
\begin{align}
y=e^{\pi p_zz/2}ze^{-\pi p_zz/2}=-iz,~~q=e^{\pi p_zz/2}p_ze^{-\pi p_zz/2}=
ip_z,~~ [y,q]=i,~~ [x,p]=i,
\label{2.15}
\end{align}
\begin{align}
e^{\pi p_zz/2}H_{\rm PU}e^{-\pi p_zz/2}=\bar{H}=\frac{p^2}{2}-iqx+\frac{1}{2}\left(\omega_1^2+\omega_2^2
\right)x^2+\frac{1}{2}\omega_1^2\omega_2^2y^2.
\label{2.16}
\end{align}
With its factor of $i$ the Hamiltonian $\bar{H}$ is now manifestly not Hermitian, but all of its eigenvalues are still real since they cannot change under a similarity transformation. $\bar{H}$ thus falls into the class of non-Hermitian Hamiltonians that have a $PT$ symmetry ($x$ and $y$ are $PT$ odd and $p$ and $q$ are $PT$ even) and have all energy eigenvalues real. Moreover, with $-\tfrac{1}{2}\omega_1^2\omega_2^2z^2$ being replaced by $+\tfrac{1}{2}\omega_1^2\omega_2^2y^2$ there no longer is any Ostrogradski instability. Thus all that remains is to deal with the ghosts.

However, first we must ask how $z$ could not possibly be Hermitian. It certainly is Hermitian when acting on its own eigenstates. However, that does not make it Hermitian when acting on the eigenstates of the Hamiltonian, with it instead being $y=-iz$ that is. This is the secret of $PT$ theory. With $z$ not being Hermitian, then despite the dagger notation we cannot identify $a_2^{\dagger}$ in (\ref{2.11}) as being the Hermitian conjugate of $a_2$. In fact we could have anticipated that this must be the case since otherwise $\langle\Omega|a_2a_2^\dag|\Omega\rangle$ would automatically have been positive rather than of the negative value we found for it above.

To make it manifest that all the eigenstates have positive norm we make the additional similarity transformation with a Hermitian operator $Q$ \cite{Bender2008a}
\begin{eqnarray}
e^{-Q/2}\bar{H}e^{Q/2}&=&\bar{H}^{\prime}
=\frac{p^2}{2}+\frac{q^2}{2\omega_1^2}+
\frac{1}{2}\omega_1^2x^2+\frac{1}{2}\omega_1^2\omega_2^2y^2,
\nonumber\\
Q&=&\alpha pq+\alpha\omega_1^2\omega_2^2 xy=Q^{\dagger},\qquad \alpha=\frac{1}{\omega_1\omega_2}{\rm log}\left(\frac{\omega_1+\omega_2}{\omega_1-\omega_2}\right).
\label{2.21}
\end{eqnarray}
We recognize $\bar{H}^{\prime}$ as being a fully acceptable standard, positive norm  two-dimensional oscillator system. In addition we note that with its phase being $-Q/2$ rather than $-iQ/2$, the $e^{-Q/2}$ operator is not unitary. The transformation from $\bar{H}$ to $\bar{H}^{\prime}$ is thus not a unitary transformation, but is a transformation from a skew basis with eigenvectors $|n\rangle$ to an orthogonal basis with eigenvectors 
$$|n^{\prime}\rangle=e^{-Q/2}|n\rangle,~~~~\langle n^{\prime}|=\langle n|e^{-Q/2}.$$ 
Then since $\langle n^{\prime}|m^{\prime}\rangle =\delta_{mn}$, the eigenstates of $\bar{H}$ obey
\begin{align}
&\langle n|e^{-Q}|m\rangle=\delta_{mn},\quad \sum_n|n\rangle\langle n|e^{-Q}=I,
\nonumber\\
& \bar{H}=\sum _n|n\rangle E_n\langle n|e^{-Q}, \quad
\bar{H}|n\rangle=E_n|n\rangle,\quad \langle n|e^{-Q}\bar{H}=\langle n|e^{-Q}E_n.
\label{2.23}
\end{align}
We thus recognize the inner product as being not $\langle n|m\rangle$ but $\langle n|e^{-Q}|m\rangle$, with the conjugate of $|n\rangle$ being the left-eigenvector $\langle n|e^{-Q}$. This state is also the $PT$ conjugate of $|n\rangle$, so that the inner product is the overlap of a state with its $PT$ conjugate rather than that with its Hermitian conjugate, just as we had noted earlier. And as such this inner product is positive definite since $\langle n^{\prime}|m^{\prime}\rangle =\delta_{mn}$ is. The $PU$ oscillator theory and accordingly the Pauli-Villars propagator theory are thus fully viable unitary theories.

Finally, we note that when $\omega_1=\omega_2$ the $Q$ operator becomes undefined.

\section {But where did the minus sign in the propagator go?}

The $D(k,4)$ propagator given in (\ref{2.6}) has a relative minus sign, so where is it if all norms are positive? The answer is that one should not identify the c-number $D(k,4)$ with the matrix element $\langle\Omega|T[\phi(x)\phi(0)]|\Omega\rangle$, but instead with the matrix element 
\begin{eqnarray}
D(k,4)=\int d^4x e^{ik\cdot x}\langle\Omega|e^{-Q}T[\phi(x)\phi(0)]|\Omega\rangle.
\label{1.1abc}
\end{eqnarray}
Now one can insert $\sum |n\rangle\langle n|e^{-Q}=I$ into $\langle\Omega|e^{-Q}T[\phi(x)\phi(0)]|\Omega\rangle$ and generate $D(k,4)$ with it being the introduction of $e^{-Q}$ that generates the minus sign \cite{Bender2008b} and not the presence of negative norm states. 
The error was thus in associating $\langle\Omega|T[\phi(x)\phi(0)]|\Omega\rangle$ with $D(k,4)$ without first having constructed the Hilbert space. Thus we do not actually get rid of the ghost, we show that it was not there in the first place, with the reasoning that led one to think that there is a ghost being faulty. Consequently, fourth-order derivative quantum gravity theories are not only unitary, \textbf{they always have been}.

\section{Pure fourth-order quantum gravity}

While we have established the unitarity of the second-order plus fourth-order theory associated with the generic $I_S$ given in (\ref{2.4}), from this we cannot immediately conclude by taking the limit that pure fourth-order theories such as conformal gravity with its $1/k^4$ propagator are unitary too since the limit is singular. Specifically, with $I_S$ of (\ref{2.4}) reducing to the pure fourth-order $I_S=\tfrac{1}{2}\int d^4x\partial_{\mu}\partial_{\nu}\phi\partial^{\mu}\partial^{\nu}\phi$ when $M_1^2=M_2^2=0$, in this same limit the decomposition of $D(k,4)$ into partial fractions given in (\ref{2.6}) becomes unviable. Since we set  $\omega_1=(\bar{k}^2+M_1^2)^{1/2}$, $\omega_2=(\bar{k}^2+M_2^2)^{1/2}$, in this limit $\omega_1$ and $\omega_2$ become equal, with both the commutation algebra given in (\ref{2.12}) and the $Q$ operator given in (\ref{2.21}) then becoming singular. The pure fourth-order theory thus has to be analyzed independently. 

Then,  as noted in \cite{Bender2008b}, with $Q$ being singular in the equal frequency limit the $\bar{H}$ Hamiltonian given in (\ref{2.16}) cannot be diagonalized. The equal frequency $\bar{H}$, and consequently  its field theory generalizations, becomes a Jordan--block Hamiltonian that cannot be diagonalized because it does does not have a complete set of energy eigenstates.

\section{In the Jordan-block limit where did the other eigenvector go?}

To see what happens to the eigenstates in the one-particle sector, we set $\omega_1=\omega+\epsilon$, $\omega_2=\omega-\epsilon$. In the limit $\epsilon\rightarrow 0$ we obtain 
\begin{eqnarray}
e^{i\omega_1 t}\rightarrow e^{i\omega t},\quad e^{i\omega_2 t}\rightarrow e^{i\omega t}.
\label{11.1}
\end{eqnarray}
Thus both wave functions collapse onto the same wave function, and so we lose an eigenstate. Now consider a different combination, viz. 
\begin{eqnarray}
\frac{e^{i\omega_1 t}-e^{i\omega_2 t}}{2\epsilon}\rightarrow it e^{i\omega t}.
\label{11.2}
\end{eqnarray}
This second combination has a well-defined, non-singular limit, and in the limit becomes \textbf{non-stationary}. So it is no longer an energy eigenstate. Since the two one-particle states were orthogonal to each other before we took the limit, and since they become the same eigenvector in the limit, in the limit the surviving eigenvector is both parallel to and orthogonal to itself. It  is thus has \textbf{zero norm}. Thus unlike the unequal frequency limit were all  ($PT$) norms are positive, in the limit the norms become zero.  However the non-stationary  states still stay in the Hilbert space, and the set of stationary plus non-stationary states combined is complete, so the theory is still unitary with probability still being conserved \cite{Bender2008b}. Because the eigenstates do have a non-standard but fully acceptable zero norm, terms in the probability that grow linearly in $t$ are multiplied by  zero coefficients, so that probabilities are indeed  time independent. This then is how probability conservation is maintained in the Jordan-block case, with it being completeness of the Hilbert space vectors and not completeness of the energy eigenstates that is key.

These remarks carry over to  the pure fourth-order conformal gravity theory. Thus it is a fully viable, probability conserving, non-diagonalizable, $PT$ and $CPT$ symmetric theory ($g_{\mu\nu}$ is $C$ even). The graviton has zero norm and is thus not an observable on-shell state. There is still gravitational radiation since in any covariant gravitational theory information is communicated with finite velocity (by off-shell gravitons).

\section{Conclusions}

The $PT$ option for quantum theory is now well established. It in no way changes quantum mechanics. It simply takes advantage of an option that was there from the beginning but had been overlooked. 
$PT$ symmetry is quite ubiquitous in physics, and cannot be avoided or ignored, especially when the standard $SU(3)\times SU(2)\times U(1)$ model is coupled to gravity. To establish $PT$  or $CPT$ symmetry only requires complex Lorentz invariance and probability conservation. And if the theory of quantum gravity turns out to be conformal gravity, then one of the four fundamental forces in nature would be a non-Hermitian but $PT$ symmetric theory.


\newpage

\begin{thebibliography}{99}

\bibitem{Stelle1977}Stelle K. S.,  \textit{Phys. Rev. D,} \href{https://doi.org/10.1103/PhysRevD.16.953}{\textbf{16} (1977) 953;}  \textit{Gen. Relativ. Gravit.,} \href{https://doi.org/10.1007/BF00760427}{\textbf{9} (1978) 353.}



\bibitem{Pauli1949} Pauli W. and Villars F.,  \textit{Rev. Mod. Phys.,} \href{https://doi.org/10.1103/RevModPhys.21.434}{\textbf{21} (1949)  434.} 

\bibitem{Pais1950}Pais A. and Uhlenbeck G. E.,  \textit{Phys. Rev.,}  \href{https://doi.org/10.1103/PhysRev.79.145} {\textbf{79} (1950) 145.}

\bibitem{Mannheim2017} Mannheim P. D.,  \textit{Prog. Part. Nucl. Phys.,} \href{https://doi.org/10.1016/j.ppnp.2017.02.001} {\textbf{94} (2017) 125.} 

\bibitem{tHooft2015b} 't Hooft G., Singularities, horizons, firewalls, and local conformal symmetry \textit{Preprint} \href{https://arxiv.org/abs/1511.04427}{arXiv:1511.04427 [gr-qc]} (2015).

\bibitem{tHooft2010a} 't Hooft G.,  Probing the small distance structure of canonical quantum gravity using the conformal group \textit{Preprint}  \href{https://arxiv.org/abs/1009.0669}{arXiv:1009.0669 [gr-qc]} (2010).

\bibitem{Mannheim2006} Mannheim P. D.,  \textit{Prog. Part. Nucl. Phys.,}  \href{https://doi.org/10.1016/j.ppnp.2005.08.001}{\textbf{56} (2006) 340.} 

\bibitem{Bender2008a}Bender C. M. and Mannheim P. D.,   \textit{Phys. Rev. Lett.,}  \href{https://doi.org/10.1103/PhysRevLett.100.110402}{\textbf{100} (2008) 110402.}

\bibitem{Bender2008b}Bender C. M. and Mannheim P. D.,    \textit{Phys. Rev. D,}  \href{https://doi.org/10.1103/PhysRevD.78.025022}{\textbf{78} (2008) 025022.}

\bibitem{Bender2018} Bender C. M., \textit{$PT$ Symmetry in Quantum and Classical Physics} (Singapore: World Scientific Press) 2018.

\bibitem{Mannheim2018a}  Mannheim P. D., \textit{J. Phys. A: Math. Theor.,}  \href{https://doi.org/10.1088/1751-8121/aac035}{\textbf{51} (2018) 315302.}

 \bibitem{Mannheim2018c} Mannheim P. D.,  \textit{ Phys. Rev. D,}  \href{https://doi.org/10.1103/PhysRevD.97.045001}{\textbf{97} (2018) 045001.}


\bibitem{Ostrogradski1850}  Ostrogradski M., \textit{Memoires sur les equations differentielles relatives au probleme des isoperimetres}  \textit{Mem. Ac. St. Petersbourg, VI,} \textbf{4} (1850) 385.

\bibitem{Mannheim2005} Mannheim P. D. and  Davidson A., 2000 \textit{Preprint} \href{https://arxiv.org/abs/hep-th/0001115}{ hep-th/0001115;}  \textit{Phys. Rev. A,} \href{https://doi.org/10.1103/PhysRevA.71.042110}{\textbf{71} (2005) 042110.} 

\bibitem{Lee1954} Lee T. D.,  \textit{Phys. Rev.,} \href{https://doi.org/10.1103/PhysRev.95.1329}{ \textbf{95} (1954) 1329.}


\bibitem{Kallen1955}K\"all\'en G. and  Pauli W.,   \textit{Mat. Fys. Medd. Dan. Vid. 
Selsk.,} \href{https://doi.org/10.1007/978-3-319-00627-7_94} {\textbf{30} (1955) 7.}

\bibitem{Heisenberg1957} Heisenberg W.,  \textit{Nucl. Phys.,} \href{https://doi.org/10.1016/0029-5582(87)90060-5}{\textbf{4} (1957) 532.}




\bibitem{Bender2005} Bender C. M., Brandt S. F., Chen J.-H. and Wang Q.,  \textit{Phys. Rev. D,} \href{https://doi.org/10.1103/PhysRevD.71.025014}{\textbf{71} (2005) 025014.}





\end{thebibliography}
\end{document}